\documentclass{sf2a-conf2023}
\usepackage{graphicx}
\usepackage{hyperref}
\usepackage[]{natbib}  
\usepackage{epstopdf}

\def\BibTeX{{\rm B\kern-.05em{\sc i\kern-.025em b}\kern-.08em
    T\kern-.1667em\lower.7ex\hbox{E}\kern-.125emX}}
\bibpunct{(}{)}{;}{a}{}{,}  

\begin{document}

\TitreGlobal{SF2A 2023}


\title{Multi-wavelength studies on Fast Radio Bursts}

\runningtitle{Multi-wavelength studies on FRBs}

\author{C. NG}\address{Laboratoire de Physique et Chimie de l'Environnement et de l'Espace - Université d'Orléans/CNRS, 45071, Orléans Cedex 02, France}

\setcounter{page}{1}


\maketitle


\begin{abstract}
Multi-wavelength (MW) observations of Fast Radio Bursts (FRBs) is a key avenue to uncover the yet-unknown origin(s) of these extragalactic signals. In this proceeding, we discuss the need for precise localization to conduct MW studies. We present a number of theoretical predictions of MW counterparts and mention a few examples of on-going MW campaigns. 
\end{abstract}



\section{Introduction}
FRBs are short ($\mu$s to ms), bright ($<$10$^{44}$\,erg\,s$^{-1}$) bursts from extra-galactic distances ($z\sim0.03$ to $z>1$). 
They can potentially be used as cosmological probes, for example, to study the intergalactic medium \citep{ravi2016} and find the missing baryons \citep{Macquart2020}. 
See \citet{Ng2023} for a brief review of the current status of the FRB field. 

Since the discovery of the first FRB in 2007 \citep{Lorimer2007}, the origin(s) of FRBs remains a highly contentious topic. MW observations is arguably the only unambiguous way to solve this puzzle. 
Despite a number of MW efforts in the literature, FRBs have thus far only been detected between 110\,MHz \citep{Pleunis2021lofar} and 8\,GHz \citep{Zhang2018}.
There is not yet any definitive detections at wavelengths other than radio, except the FRB-like signal detected from the Galactic magnetar SGR~1935+2154 \citep{SGR-CHIME,Bochenek2020,Mereghetti2020}.
To-date, about 40 FRBs\footnote{See the Transient Name Server (TNS) database \url{https://www.wis-tns.org/search} for up-to-date numbers.} have been localized to a host galaxy, showing diverse host galaxy properties and local environments \citep{Gordon2023,Mannings2021}.



\begin{figure}[ht!]
  \vspace{-0.1cm}
 \centering
\includegraphics[width=0.73\textwidth,clip]{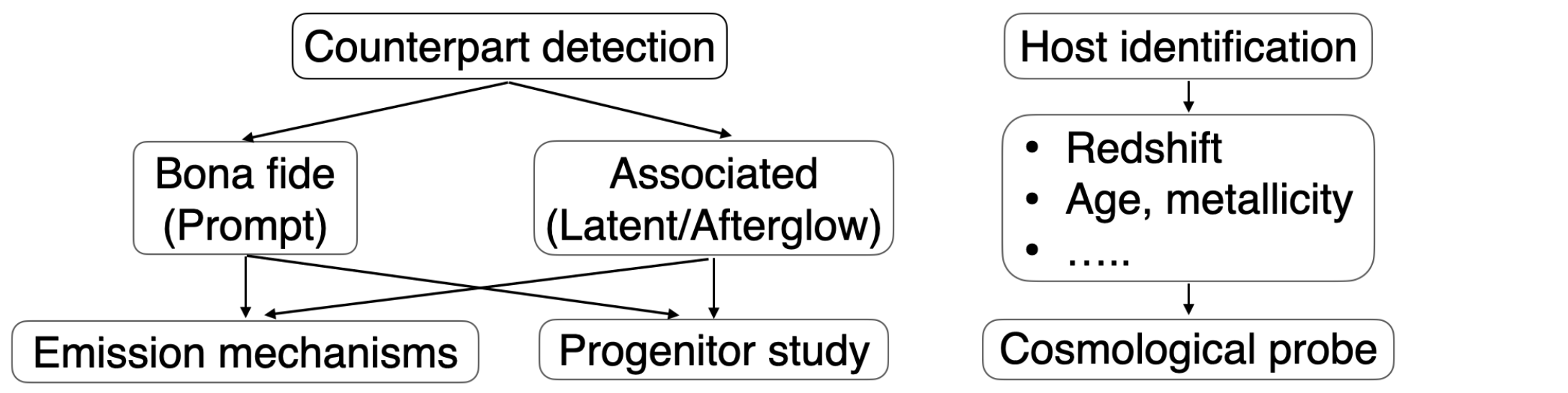}    
  \vspace{-0.3cm}
  \caption{A schematic of the various types of MW follow-up discussed in this proceeding.}
  \label{fig:schema}

\end{figure}

Fig.~\ref{fig:schema} attempts to summarize various types of MW studies, with the two main motivations being the detection of counterparts and the identification of host galaxies. There are two categories of counterparts:
\begin{itemize}
    \item Prompt / bona fide: Counterpart just before or near the same time as the FRB.
    \item Associated / latent / afterglow: Counterpart well before or after the FRB, typically associated to the environment, or source, and does not have to be with the exact burst.
\end{itemize}
A positive detection of either bona-fide or associated counterparts can help set constraints on the model of the emission mechanisms and tell us what the progenitor might be. 
For host identification, the goal is essentially to locate the FRB to a host galaxy, derive the redshift, spectral properties, age, metallicity and other relevant information. Apart from providing us with a better understanding of the FRB environment, these information are also crucial for conducting cosmological studies using FRBs as probes.

\section{The need for localization}
Before doing any multi-wavelength study, we need to have a precise idea of where the FRB is located. 
Most standalone radio telescopes do not have good enough localization capability to tell which host galaxy the FRB came from, except for the closest sources.
Take the CHIME radio telescope as an example. 
When it first detected the Galactic magnetar SGR~1935+2154, the localization uncertainty was large (about 1\,deg) as it was seen from the sidelobe \citep{SGR-CHIME}. 
Fortunately, the relatively low Dispersion Measure (DM) of this source set tight constraints on the distance. 
With the same two bursts detected by X-ray telescopes \citep{Mereghetti2020}, it was clear that they are from the same source. 
For the closest $\sim$1\% of CHIME FRB discoveries, it is possible to use their low DMs to set upper limit to the redshifts, reduce chance coincidence and in turn, pinpoint the host galaxy. 
This was the case for the first publication of the M81 FRB \citep{Bhardwaj2021}, at least before the European VLBI Network (EVN) looked at it and further improved the localization \citep{Kirsten2022,Marcote2022}. 
Note that closer by FRBs also tend to be brighter (although not necessarily), which means the localization uncertainties can be reduced thanks to the high signal-to-noise detections. 

To achieve proper localization we will need interferometric telescopes, which can provide precision of around 0.1\,arcsec and allow a confident association of host galaxy \citep[see, e.g. the analysis conducted with the Very Large Array (VLA)][]{Chatterjee2017}.
With a larger scale interferometer such as the EVN, we can have of the order of 1\,mas localization capability, meaning we can tell exactly which part of the host galaxy did the FRB originate \citep{Kirsten2022}. 
However, the EVN has a small field-of-view and coordinating observations with it is an intensive operation. Its time allocation is heavily subscribed and we will definitely not be able to use EVN for every single FRB.
On the positive side, a number of next generation radio facilities are coming online.
CHIME is commissioning two outrigger stations located in Hat Creek ($\sim$1000\,km baseline from CHIME) and Green Bank ($\sim$3000\,km baseline) in the US. 
Once operational, CHIME will be able to achieve $\sim$100\,mas-localization for all its FRBs.
In the near future, 
CHORD \citep{CHORD}, DSA-2000 \citep{DSA2000}, the Square Kilometre Array (SKA) and BURSTT \citep{Lin2022} will provide of the order of 500 FRBs per month each with milliarcsecond-level localization uncertainty.

\section{Theoretical counterpart expectations}
Theorists have predicted a number of bona-fide and associated counterparts for FRBs. Here we will discuss a few examples:

\subsection{Bona-fide counterparts}
There are two main models for prompt counterparts. In the synchrotron maser emission model, relativistic ejecta runs into a slower moving medium. The flare collides with an ion shell and the shell decelerates through shock waves which gives rise to an FRB burst \citep[see, e.g.][]{Metzger2019}. 
In this model, we would also get incoherent emission in hard energy (gamma ray and X-ray).
The other model involves a classic Magnetar flare, where the trapped fireball also emits thermal X-ray and comptonization \citep[see, e.g.][]{thompson1995,Lu2020}. 
The Galactic magnetar SGR~1935+2154 might be an example of this. 

\subsection{Associated counterparts}
A wide range of possibilities have been proposed in the literature, spanning pretty much the entire spectrum from radio, optical, X-ray, gamma ray to even gravitational waves (GW):

\begin{itemize}
    \item Radio: Persistent radio source (PRS) associated with a magnetar pulsar nebula has been suggested. This is particularly relevant given a few FRBs have been found to co-locate with PRS \citep{Law2022}. 
    \item Optical: If FRBs are associated with neutron stars, we might expect associated supernova explosions some time in the past since neutron stars are primarily born in core-collapse supernovae \citep{Margalit2018}. Since the environment around the progenitor is initially very dense, we would probably have to wait at least a decade or so between the initial supernova and the time when the environment has dissipated enough for an FRB to be seen. One could look into archival data for a supernova association, but the risk of getting false chance associations could be quite high. 
    \item X-ray: Synchrotron maser emission from persistent X-ray / Ultra Luminous X-ray Sources (ULX)-like objects has been suggested \citep{Sridhar2021b,Deng2021}. This is probably going to be difficult to detect given the extragalactic distances of FRBs. One would need very sensitive X-ray telescopes and detections might only be feasible for the brightest sources.  
    \item Gravitational waves (GW) / Gamma ray: Short gamma ray burst (GBR) from neutron star mergers is a possible albeit rare formation channel for magnetars \citep{Margalit2019,Wang2020,Sridhar2021}. FRB can be formed from the interaction of the magnetospheres of the neutron stars as they inspiral towards each other. It has been proposed that in this model, we could get an FRB preceding an associated GW event.
    Alternatively, if a stable, long-lived neutron star is formed after the merger, an FRB can be produced through the kilonova ejecta. In this case we will need to wait a few weeks or so for the FRB to go through this less dense, rapidly expanding media.
\end{itemize}

\section{Examples of MW campaigns}
MW campaigns are logistically challenging and require coordination from the community. The detection of bona-fide counterparts is particularly tricky because time is critical. One has to get on source basically in real-time of the FRB burst. On the other hand, associated counterpart detections have less of a rush as emission is expected on longer timescales. There are a number of approaches in the literature:
\begin{itemize}
    \item Take a radio telescope as the master and employ facilities at other wavelengths to shadow it, for example in the case of MeerKAT+MeerLITCT \citep{MeerLITCH}. It is probably a good idea to pick a radio telescope that has already seen FRBs in the past, to have a better chance of success!
    \item A reversed search, where a radio telescope is used to to shadow, for example, Swift/XRT \citep[see, e.g.][]{Nicastro2021}. This is probably applicable for smaller, less heavily subscribed radio telescopes. 
    \item A large-scale campaign coordinating many telescopes across the spectrum to observe the same program together. A great example is the Deeper Wider Faster (DWF)\footnote{DWF: \url{https://www.swinburne.edu.au/research/centres-groups-clinics/centre-for-astrophysics-supercomputing/our-research/data-intensive-astronomy-software-instrumentation/deeper-wider-faster-program/}} project.
    \item Search archival data of other wavelengths at the positions of known FRBs to see if there is any associated counterparts from further back in time \citep[see, e.g.][]{Ashkar2023}. 
\end{itemize}

In order to facilitate these time-critical follow-up studies, CHIME publishes quasi-real time alerts for FRB discoveries via the framework of VOEvents\footnote{CHIME VOEvents: \url{https://www.chime-frb.ca/voevents}} on a (free) subscription basis. 
As the CHIME real-time FRB pipeline detects a candidate, it will be sent to a database and the VOEvent will be published by the Comet broker, which is an implementation of the VOEvent The Transport Protocol broadcasts machine-reachable messages over the Internet to any observatories or send out human-readable emails.
The end-to-end processing is very fast. VOEvents can be received by other telescopes within about 30\,seconds of an FRB detection in CHIME.
This service has been available since October 2021. So far, over 1800 alerts have been published to over 100 subscribers from over 60 institutes. 

For the study of host galaxy identification and classification, it typically involves conducting coordinated photometric and spectroscopic follow-up observations. 
This type of follow up also has no rush but obtaining observing time on telescopes sensitive enough to observe the host galaxy is not always for granted. 
So far, most of these follow-up studies are conducted on a case by case basis, targeting individual FRB one at a time. 
Nonetheless, a more coordinated effort has been initiated by the Fast and Fortunate for FRB Follow-up (F$^{4}$)\footnote{F$^{4}$: \url{https://sites.google.com/ucolick.org/f-4}} collaboration. They have signed agreement with a number of FRB discovery radio experiments (e.g. CHIME, CRAFT, AstroFlash, MeerTRAP), to take all their arcsecond-localized FRBs and go to telescopes at other wavelength to identify the hosts. They have partnerships with the Keck, Gemini, Magellan, ESO/VLT, Hubble, ALMA, Chandra, and the VLA. 

\section{Conclusion}
Multi-wavelength observation is arguably the only unambiguous way to identify counterparts for FRBs.
So far FRBs have only been detected in radio (110\,MHz--8\,GHz), except one FRB-like signal from the Galactic magnetar SGR~1935+2154.
Multi-wavelength observations require coordinated community effort, and a number of on-going campaigns are summarized in this proceeding. 
Low-DM FRBs have a better chance of having multi-wavelength detections and are possibly worth prioritizing for follow-up studies. 
We will hopefully soon arrive at the era of many (well-)localized FRB events, and at which point, 
there will also be potential synergies with publicly available catalogues, for example, PanSTARRS, SDSS, DESI, Rubin/LSST, Euclid, SPHEREx, etc, for host galaxy studies.


\begin{acknowledgements}
\end{acknowledgements}

\bibliographystyle{aa}  
\bibliography{NG-GUIHENEUF2} 

\end{document}